\documentstyle[preprint,aps,prl,epsf]{revtex}

\begin{document}

\title{{\bf Molecular dynamics of flows in the Knudsen regime}}

\author{{\sc Marek Cieplak$^{1}$, Joel Koplik$^{2}$ 
\& Jayanth R. Banavar$^{3}$ }}

\address{$^{1}$ Institue of Physics, Polish Academy of Sciences,
02-668 Warsaw, Poland} 

\address{$^{2}$ Benjamin Levich Institute and Department of Physics,
City College of New York, NY 10031} 

\address{$^{3}$ Department of Physics and Center for Materials Physics,
104 Davey Laboratory, The Pennsylvania State University, University
Park, Pennsylvania 16802}

\address{ $\;\;$}

\address{ $\;\;$}

\address{
\centering{
\medskip\em
{}~\\
\begin{minipage}{14cm}
Novel technological applications
often involve fluid flows in the Knudsen regime in which
the mean free path is comparable to the system size.
We use molecular dynamics simulations to study the transition
between the dilute gas and the dense fluid regimes
as the fluid density is increased.
{}~\\
{}~
\end{minipage}
}}

\maketitle

\newpage 

In a landmark paper \cite{1}, Maxwell applied his kinetic 
theory \cite{2} to the
study of the slip length associated with fluid flow close to
a solid surface. He postulated a linear combination of
two kinds of behaviour when a 
molecule was incident on a wall -- it is either specularly 
reflected or it 
undergoes enough collisions with the wall molecules so that it loses
memory of the incident velocity and emerges with a velocity characteristic
of the wall temperature.
We revisit this problem with molecular
dynamics simulations -- as the properties of the wall are varied,
we find both kinds of behaviour suggested by Maxwell, but the
crossover from one to the other is not as simple as
envisioned by him. In addition, rich behaviour
is found in the fluid flow properties on varying the fluid density
from the dense liquid regime to
the Knudsen regime (large mean free path
compared to the system size).
Our studies of flows in the subcontinuum regime provide a unique
window on the nature of boundary conditions at a fluid - solid
interface and are relevant to 
the behaviour of nanoscale mechanical devices,
high altitude aerodynamics, aerosol dynamics, isotope separation
schemes, flow cooling techniques, and energy transfer in
molecular collisions \cite{3,4,5,6}.

Studies of flows at nanoscales  involve questions
pertaining to a sub-continuum regime \cite{7,8} for
which experiments are often unable to provide the requisite answers.
The behaviour of materials at the nanoscale is often
quite different from that in the bulk \cite{9,10,11,12}.
Numerical simulations based on
molecular dynamics (MD) offer a natural tool to handle
the interacting many-body nanoscale physics.
A detailed understanding of the behaviour of materials at very small
scales is necessary for the fabrication of miniature devices and for
the manipulation of materials at the molecular scale.  
Such microscopic studies
are useful for assessing the range of validity of traditional continuum
theories. In addition,  
atomistic simulations are an invaluable tool for deducing
constitutive relationships, which may then be fed into continuum
calculations.  

Here, we report on MD studies of the boundary conditions
in single-fluid flows in the Knudsen regime \cite{13}
in which the mean free path, $\lambda$,  becomes comparable to the
characteristic device dimension, $L$.
Such flows behave very differently from bulk, continuum expectations.
There is considerable work on the
modified boundary conditions
based on either semi-heuristic kinetic theory arguments
or using an analysis of the Boltzmann equation, but inevitably
severe simplifications are made concerning the dynamics
of the system \cite{14}. Furthermore, in such treatments,
the solid boundaries are not handled in a realistic
manner as having atomic structure on length scales
comparable to that of the fluid constitutents \cite{15,16}.

We study Lennard-Jones fluids
undergoing two kinds of flow, Couette and Poiseuille, between two
solid walls with an atomic structure. Three kinds of walls,
two attractive and one repulsive, are discussed.
We  consider a range of fluid densities to study the crossover from
flows described by continuum fluid mechanics to dilute gas flows
in the Knudsen regime.

In a Lennard-Jones
fluid, two atoms separated by a distance $r$
interact with the potential
\begin{equation}
V_{LJ}(r)\;=\; 4 \epsilon
\left[ (\frac{r}{\sigma})^{-12} \;-\;
(\frac{r}{\sigma})^{-6} \right] \;\;,
\end{equation}
where $\sigma$ is the
size of the repulsive core
[the potential is truncated at $2.2 \sigma$].
The flows take place between two solid planar
walls parallel to the $xy$ plane with periodic boundary conditions
imposed along the $x$ and $y$ directions.
Couette flow was generated by moving the walls at constant velocity 
in opposite directions along the $x$-axis
whereas Poiseuille flow was induced by an application
of uniform acceleration $g$ on all the fluid molecules in the $x$-direction.

Following Thompson and Robbins \cite{17},
the wall of atoms form two [001]
planes of an fcc lattice with
each wall atom tethered to a fixed lattice site by a harmonic
spring of spring constant $k$. 
Our studies where done for
$k$=400 so that the mean-square displacement about the
lattice sites did not exceed the Lindemann criterion for melting
but was close to the value set by this criterion.
The wall fluid interactions were modelled by another Lennard-Jones
potential,
\begin{equation}
V_{wf}(r)\;=\; 16 \epsilon 
\left[ (\frac{r}{\sigma})^{-12} \;-\; A \;
(\frac{r}{\sigma})^{-6} \right] \;\;,
\end{equation}
with the parameter $A$ varying between 1 and 0, corresponding to
attractive and repulsive
walls respectively. For the narrowest channel that we studied,
the fluid atoms 
were confined to a space measuring 13.6$\sigma$ by 5.1$\sigma$
in the $xy$ plane and 12.75$\sigma$ between the walls.
The ``basic'' walls in our studies
contain 96
wall atoms each (in each wall plane, the atomic periodicity is 
0.85$\sigma$). 
We have also considered the case of 
walls
which are
about 2.5 times denser and consist of 176 wall atoms each --
we shall refer to these as ``dense'' walls.

The equations of motion were integrated using a fifth-order Gear
predictor-corrector algorithm \cite{18} with a step size of $0.005\tau$,
where $\tau =(m \sigma ^2 /\epsilon)^{1/2}$ and $m$ is the atomic
(wall and fluid) mass.
The fluid and wall temperatures were fixed
at $k_BT/\epsilon$=1.1. We present results obtained with the use
of  Langevin thermostat \cite{19} but the results obtained with the use
of the Nose-Hoover thermostat \cite{20,21} were virtually identical.
The noise was applied in all directions during equilibration
and only in the $y$-direction during the data-taking phase.
The averages were obtained over runs which lasted at least 5000 $\tau$,
and in the extreme dilution case, for up to 400000 $\tau$.
An initial time of at least 400$\tau$ was spent  
for equilibration.
The spatial averaging was carried out in slabs of width $\sigma /4$
along the $z$-axis.

In the dilute gas limit,
the Knudsen number, $Kn\;$,
is defined as $\lambda /L$, i.e., as 
$\; (\sqrt{2} \pi \rho \sigma ^2 L)^{-1}$, where
$\lambda$ is the mean free path,
$\rho$ is the fluid number density and $L$ is the characteristic width
of the channel. A study of the density profile shows qualitatively
different behaviours of the density profiles for the attractive and
repulsive walls. In the former case, 
two adsorbed layers form near both walls whereas there
is merely a depletion zone, whose thickness is of order $\sigma$,
alongside the repulsive walls.
The first layer at the attractive walls is fairly periodic, as measured
by the static structure factor, and the resulting local order
corresponds to either a  square or hexagonal lattice depending on
whether the wall is basic or dense. 
$L$ is defined as
an effective channel width in which the flow takes place (in the attractive
case this is the distance between centers of the second layers).
In the center of the channel, the fluid density is essentially
constant and this value is chosen as $\rho$ in the definition of $Kn$.

Figure 1 shows the Poiseuille flow velocity profiles for three
different values of $\rho$ (or $Kn$) for the attractive wall case. 
We have confirmed with studies of
channels of three different widths and the two wall densities that
the crossover value of $\rho$ above  which continuum fluid mechanics
holds, $\rho _K$, is somewhat less than 0.19, corresponding to a $Kn$ of
order 0.1 for the channel of the smallest width. In this regime,
the profiles are clearly parabolic and essentially with no slip
at the walls. In contrast, the velocity profile at the highest $Kn$
that we were able to study (corresponding to just 4 particles in the
interior of the channel) is almost that of plug flow with
a huge slip length. 
For repulsive walls, the plug flow regime extends to higher
densities because such walls provide little coupling
to the fluid and correspondingly,
$\rho _K$ becomes of order 0.3.
Furthermore,
in the continuum regime, the velocity field exhibits substantial slip.
A consistent behaviour is found for Couette flow with higher slip
for the repulsive walls (compared to the attractive wall case)
for $\rho > \rho _K$ and essentially zero
fluid velocity at the walls at the highest $Kn$.

We now proceed to study the effect of interactions
between the fluid particles undergoing Poiseuille flow on
increasing $\rho$ within the channel. 
In Poiseuille flow, the velocity field is
highest in the middle of the channel and is denoted by $v_{\rm max}$.
At the lowest densities, one expects
that $v_{\rm max}$ ought to scale as $L$. This is because the driving
force is effective in accelerating
the fluid particles only during the time spent traversing the channel
between collisions with the walls (and the associated adsorbed layers).
This time simply scales as the channel width. On the other hand,
at large $\rho$, corresponding to usual fluid behaviour,
$v_{\rm max}$ ought to scale as $L^2$ as given by the Navier-Stokes equation
with no-slip boundary conditions.
Figure 2 shows a plot of how $v_{\rm max}$ interpolates between
the low and high Knudsen number behaviours
as the fluid density is varied. 
Strikingly, $v_{\rm max}$ shows a maximum
around $\rho _K$ with the peak being more pronounced
for larger $L$'s. We have confirmed that this qualitative
behaviour is essentially independent of the nature of the walls.
Physically, $\rho _K$ corresponds to the situation when
the fluid intermolecular interactions begin to dominate leading
to a diminished influence of the wall in the interior of the
fluid.

Figure 3 shows a vivid illustration of the role played by the
intermolecular interactions in leading to an enhancement of
$v_{\rm max}$ even as the fluid density is moderately increased from the
high $Kn$ limit. For a repulsive wall (not shown), the particle
is reflected approximately specularly and moves out of the vicinity
of the wall immediately. The bottom panel shows qualitatively different
behaviour in the (basic) attractive wall case. 
Here, again as suggested by Maxwell,
a particle either thermalizes in the vicinity of the wall for a
relatively long time and then escapes from it or occasionally still
undergoes specular reflection. On increasing the fluid density,
one gets the situation illustrated in the middle panel in which
intermolecular collisions keep the particles longer within
the channel thus reducing the time of localization near the walls and
effectively increasing the velocity field in the middle of the channel
(Figure 2). At still higher densities ($\sim \rho _K$), the intermolecular
interactions become sufficiently large so that liquid-like behaviour
sets in and $v_{\rm max}$ begins to decrease.

A key quantity that characterizes the fluid solid interface 
is the collision kernel \cite{14,22}
-- the probability density that a molecule striking the surface
with a given velocity reemerges with a specific velocity after
a given residence time. For the repulsive wall, we find that Maxwell's
specular reflection scenario holds exceedingly well and the residence
time is negligible. On the other hand, for the attractive walls
the distribution of residence times develops a long tail. Strikingly,
the velocity distribution of a particle after the collision with the 
wall is substantially independent of the incoming velocity.
As an illustration, Figure 4 shows the case of particles
with unit speed incident normal to the wall.
The inset shows the probability distribution of residence times
whereas the main figure shows the distributions of the normal ($z$)
and tangential ($x$) velocities of the scattered molecules.
The results are in excellent accord with Maxwell's thermal wall
scenario in which the probability densities corresponding to $v_x$ and $v_z$
are \cite{1,13,23}
\begin{equation}
\phi _x(v_x)=\sqrt{\frac{m}{2\pi k_BT}} \; exp (-\frac{mv_x^2}{2k_BT})
\end{equation}
and
\begin{equation}
\phi _z(v_z)=\frac{m}{k_BT} \;v_z \; exp (-\frac{mv_z^2}{2k_BT})
\end{equation}
respectively. As expected for a thermal wall, the residence time is
found to be uncorrelated with the outgoing velocity.

How does the behaviour crossover from that of specular reflection
to that of a thermal wall?
Following Maxwell$^1$, a common hypothesis
is a linear combination of both behaviours
with weights $(1-f)$ and $f$ respectively.
Such an intermediate
case is realized, e.g., 
on considering the scattering from a wall with
the wall-fluid potential 
given by Eq. 2 with $A$=1/4. 
The velocity distributions of the scattered particles
clearly deviate from pure thermal behaviour and  the data
might be interpreted as containing a specular
component. However, the resulting distributions are more complex
than the simple linear combination form \cite{24}.  The 
behaviour of the collision
kernel is further enriched on considering the role of
the interactions between the fluid molecules as the Knudsen number
decreases to a value of 0.27 corresponding to a density of
$\rho $ = 0.066 (figure 5). 
The influence of other fluid molecules leads to the distributions
more closely approaching the  thermal ones.
The nature of the boundary conditions at a fluid-solid interface is
thus complex but akin to those envisioned by Maxwell.

We are indebted to Mark Robbins and Riina Tehver for many 
valuable discussions.
This work was supported by  KBN (Grant No. 2P03B-025-13),
NASA, and the Petroleum Research Fund administered by the American
Chemical Society.

\newpage
\begin{center}
{\bf Figure captions}
\end{center} 

\vskip 1.0cm

{\bf FIG. 1}. 
Velocity profiles of Poiseuille flow with $g= 0.01 \sigma / \tau ^2$
for the smallest width channel with basic attractive walls.
The values of the interior fluid density and the corresponding
Knudsen number are indicated. The filled circles denote the highest
$Kn$ case. The inner wall layers are at the edges of the figure.
Effectively, one observes no slip
boundary conditions for the two higher density cases
(the slip length is around 1.7$\sigma$ measured with respect to the
wall location, but note the presence of the adsorbed layers at the wall).

\vskip 1.0cm

{\bf FIG. 2.} 
Plot of the fluid velocity in the middle of the channel in Poiseuille flow 
with $g=0.01 \tau /\sigma ^2$ as a function of fluid density (and
the Knudsen number in the inset). The plot is for three different
channel widths. The $L_0$ in the figure denotes the narrowest channel.
For this case, both attractive walls were considered --
the basic (hexagons) and higher (asterisks) wall densities.
$2L_0$ and $4L_0$ denote channels that are effectively twice and four times 
as wide as the smallest channel.

\vskip 1.0cm

{\bf FIG. 3.}
Plot of the time dependence of the $z$-coordinate of a typical fluid
particle undergoing Poiseuille flow in the narrowest channel. 
The horizontal dotted lines indicate the $z$-coordinates
of the wall molecules. The solid horizontal line 
refers to the centers of the adsorbed first layers.

\vskip 1.0 cm

{\bf FIG. 4.}
The probability density distributions for the $z$ (normal) and $x$
components of the outgoing velocity of a fluid molecule after
collision with the wall and the associated adsorbed layer. 
The solid histograms are obtained based on 3000 scattering events.
The dotted lines are the analytic predictions of Maxwell, eqs. 3 and 4,
discretized with a bin size of 0.1 $\sigma / \tau _0$
as in the simulations.
The inset shows the distribution of residence times.
The residence time is measured as the time interval between a
fluid molecule leaving a force-free region (near the wall) and
its reemergence. The peak of the curve (at a non-zero value of $\tau$)
corresponds to a virtually instantaneous bounce back of the fluid
molecule and accounts for the time needed to approach the wall and
return.



\begin{figure}
\vspace*{2cm}
\epsfxsize=5.4in
\centerline{\epsffile{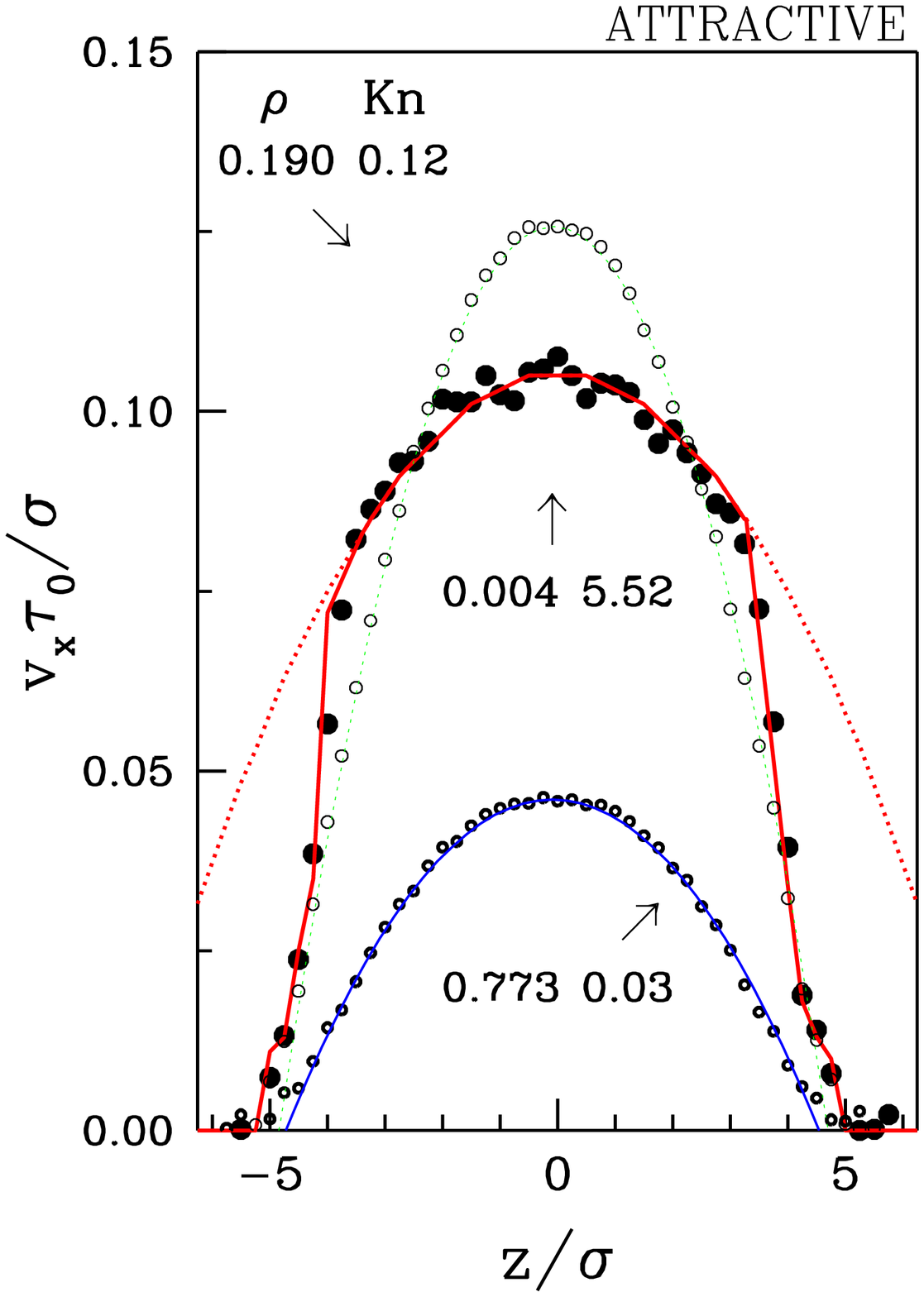}}
\vspace*{1cm}
\caption{ }
\end{figure}

\begin{figure}
\epsfxsize=5.4in
\centerline{\epsffile{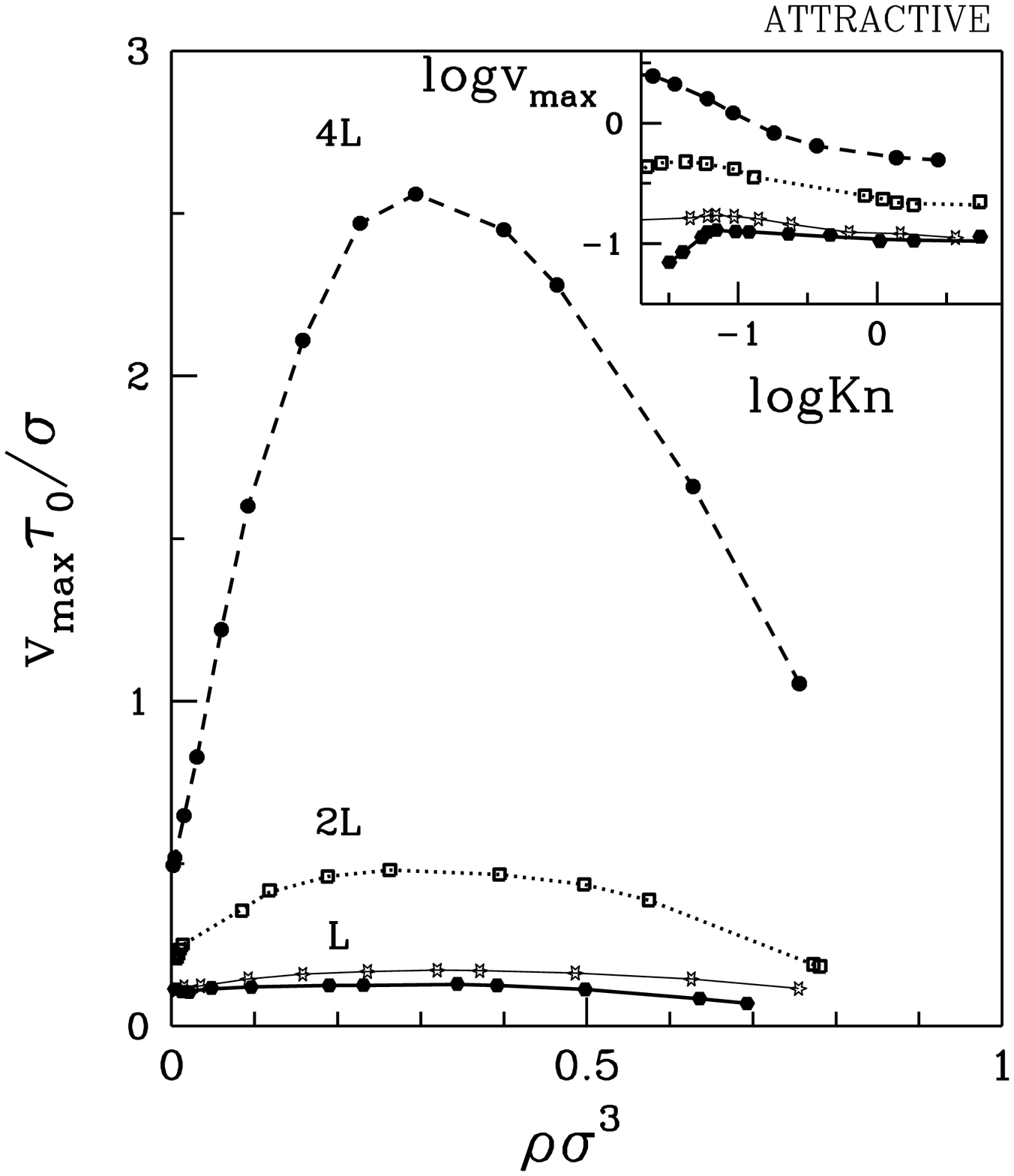}}
\vspace*{1cm}
\caption{ }
\end{figure}

\begin{figure}
\epsfxsize=5.4in
\centerline{\epsffile{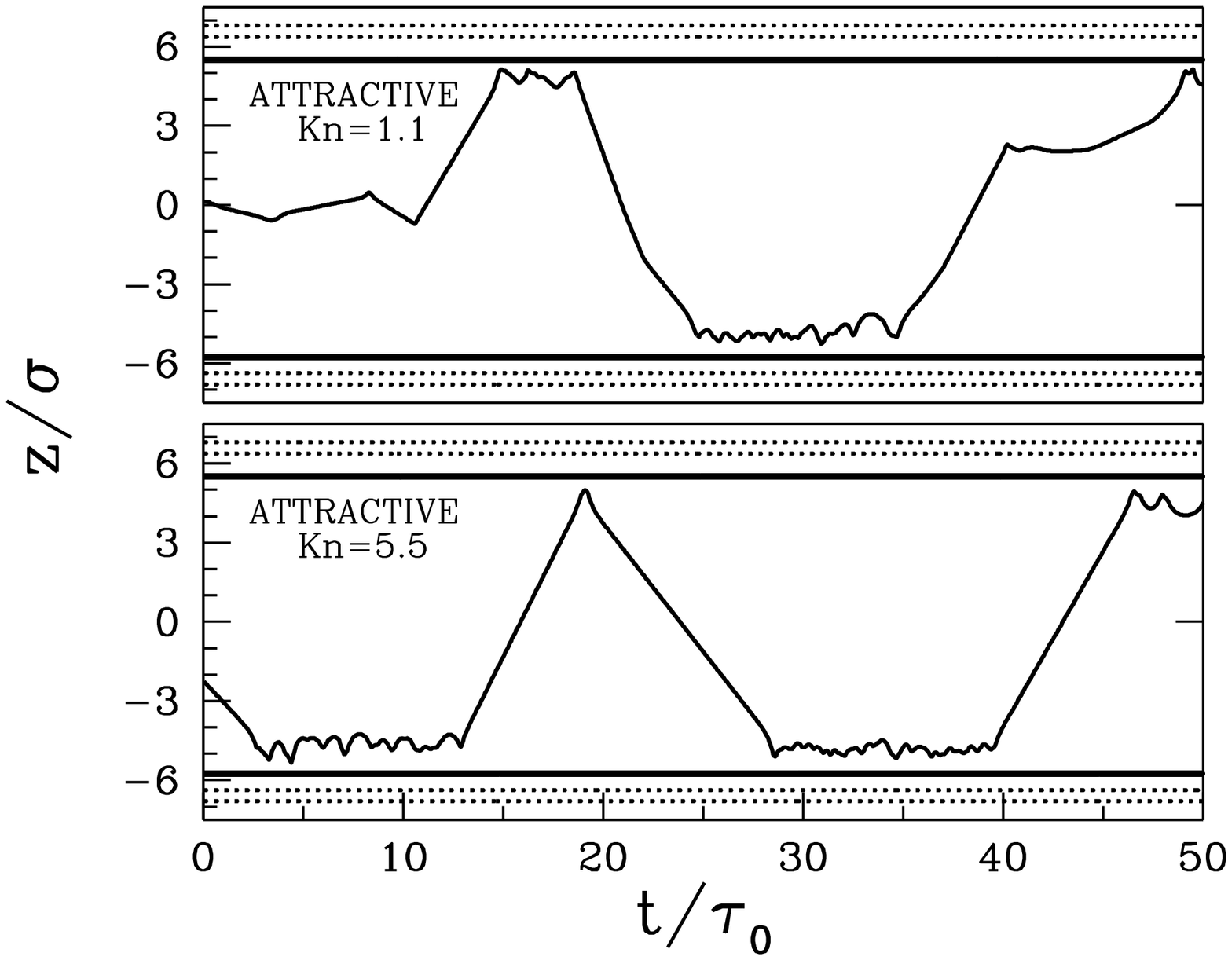}}
\vspace*{3cm}
\caption{ }
\end{figure}

\begin{figure}
\epsfxsize=5.4in
\centerline{\epsffile{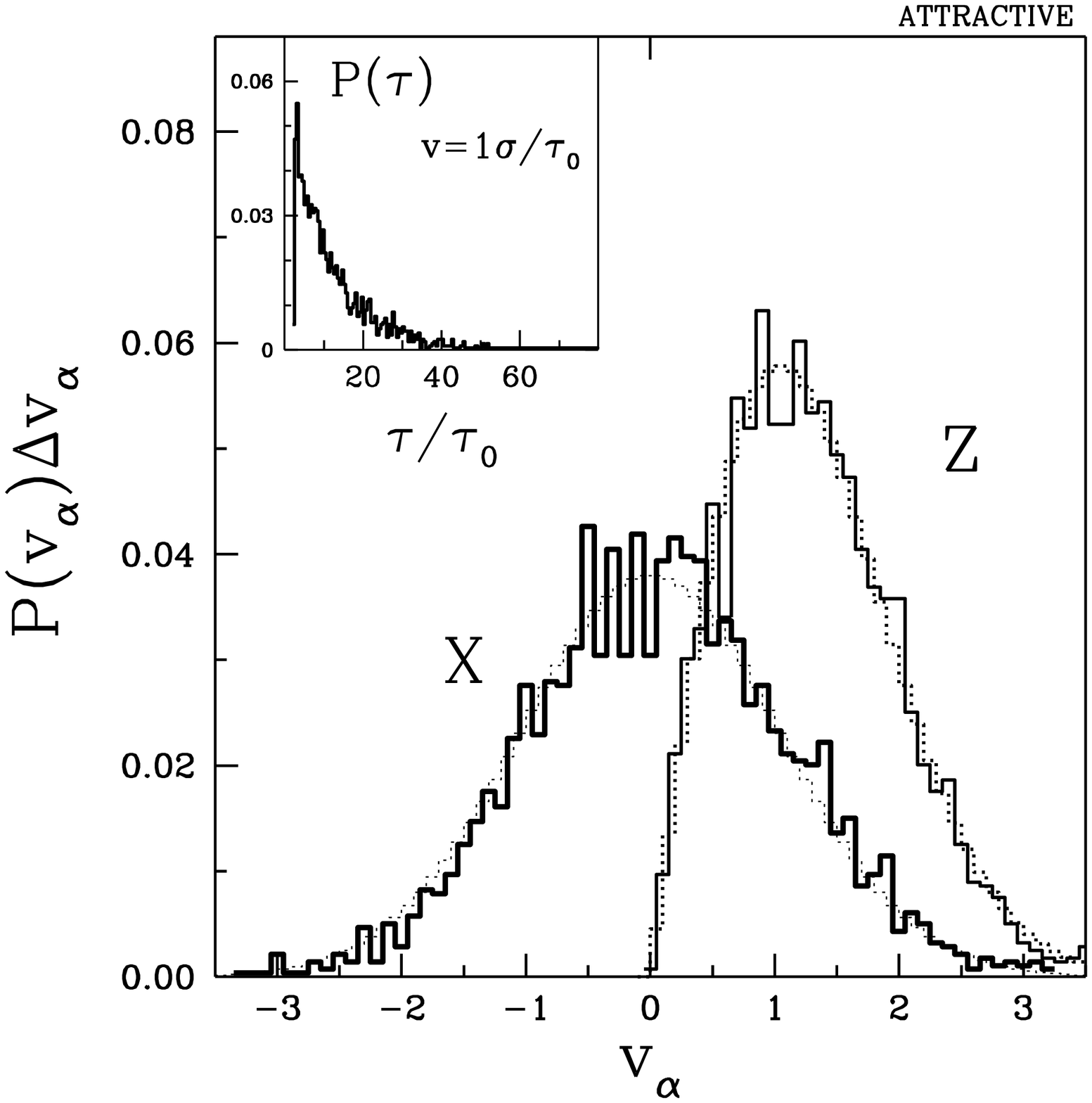}}
\vspace*{0.5cm}
\caption{ }
\end{figure}

\end{document}